# Platinum-Decorated Graphene: Experimental Insight into Growth Mechanisms and Hydrogen Adsorption Properties

Letizia Ferbel[a,+], Stefano Veronesi[a], Ylea Vlamidis[a,b], Antonio Rossi[c,d], Leonardo Sabattini[c,e], Camilla Coletti[c,d], Stefan Heun[a]

[a]NEST, Istituto Nanoscienze-CNR and Scuola Normale Superiore, Piazza San Silvestro 12, Pisa, 56127, Italy
[b]Department of Physical Science, Earth and Environment, University of Siena, Via Roma 56, Siena, 53100, Italy
[c]Center for Nanotechnology Innovation@NEST, Istituto Italiano di Tecnologia, Piazza San Silvestro 12, Pisa, 56127, Italy
[d]Graphene Laboratories, Istituto Italiano di Tecnologia, Genova, 16163, Italy
[e]NEST, Scuola Normale Superiore, Piazza San Silvestro 12, Pisa, 56126, Italy
[+] letizia.ferbel@sns.it (Corresponding author)

**Abstract.** The potential of graphene for hydrogen storage, coupled with the established role of Platinum as a catalyst for the hydrogen evolution reaction and the spillover effect, makes Pt-functionalized graphene a promising candidate for near-ambient hydrogen storage. This paper focuses on examining the process of Pt cluster formation on epitaxial graphene and assesses the suitability of the system as hydrogen storage material. Scanning tunneling microscopy unveils two primary pathways for Pt cluster growth. In the initial phase, up to ~1 ML of Pt coverage, Pt tends to randomly disperse and cover the graphene surface, while the cluster height remains essentially unchanged. Beyond a coverage of 3 ML, the nucleation of new layers on existing clusters becomes predominant. Then, the clusters mainly grow in height. Thermal desorption spectroscopy on hydrogenated Pt-decorated graphene reveals the presence of multiple hydrogen adsorption mechanisms, manifested as two Gaussian peaks superimposed on a linearly increasing background. We attribute the first peak at 150°C to hydrogen physisorbed on the surface of Pt clusters. The second peak at 430°C is attributed to chemisorption of hydrogen on the surface of the clusters, while the linearly increasing background is assigned to hydrogen bonded in the bulk of the Pt clusters. These measurements demonstrate the ability of Pt-functionalized graphene to store molecular hydrogen at temperatures that are high enough for stable hydrogen binding at room temperature.

## 1. Introduction

Hydrogen stands out as the most promising renewable energy carrier, offering a viable alternative to fossil fuels [1]. However, its widespread adoption faces several technical challenges, with storage being a significant obstacle [1-3]. To address this issue, solid state storage solutions have been explored, aiming to enable high-density hydrogen storage through physical or chemical means at near-ambient pressure and temperature [3, 4]. Graphene, with its remarkable chemical stability, lightweight nature, large surface area, and favorable physical-chemical properties for hydrogen adsorption, has emerged as a particularly captivating option. Although pristine graphene has a limited capacity for storing molecular hydrogen, especially under near-ambient conditions, the introduction of metal functionalization opens up the possibility to achieve high gravimetric densities [5-7].

Several studies, both theoretical and experimental, have examined the potential of metal functionalization of graphene for hydrogen storage using various metals. Alkaline earth metals [8, 9] and alkali metals [10-12] tend to disperse on the graphene surface or readily intercalate the graphene sheet. However, they often form metal hydrides, posing challenges for reversible hydrogen storage [13].

In contrast, transition metals (TMs) exhibit a favorable interaction known as Kubas interaction, enabling them to bind hydrogen molecules at energies conducive to room temperature storage [14-22]. TMs have also been recognized as catalysts for the dissociation of $H_2$ molecules, enabling the spillover effect when appropriately supported [23, 24]. However, TMs have a greater tendency to cluster, reducing the available active metal surface area and thereby decreasing the number of hydrogen binding sites. Consequently, in practice the storage capacity is lower than what is theoretically calculated [7, 25, 26].

Among transition metals, Platinum has long been regarded as the most efficient catalyst for the hydrogen evolution reaction and for the spillover effect [27, 28]. However, the commercial potential of this catalyst is limited by its high cost and limited availability. To address this, a promising strategy involves the use of single metal atoms or small metal clusters dispersed on supports, effectively reducing the required amount of material while maintaining the catalytic activity [29, 30]. Ab-initio calculations employing density functional theory (DFT) have demonstrated that the adsorption of Pt atoms on graphene enhances the binding energy of hydrogen molecules, from physisorption (with approximate binding energy of -0.1 eV) [6, 27, 31] in pristine graphene to chemisorption (with approximate binding energy of -0.8 eV) [31] in Pt-decorated graphene. Remarkably, it results that each Pt atom has the ability to bind up to four hydrogen molecules, in the atomic form, thereby enabling the potential for achieving high gravimetric densities at near-ambient conditions [31, 32].

The purpose of this paper is to experimentally investigate the properties and capacity of graphene functionalized with Pt for hydrogen storage applications and to verify the theoretical predictions.

## 2. Materials and Methods

Epitaxial graphene was obtained by thermal decomposition of 6H-SiC(0001) crystals at high-temperature in a BM reactor (Aixtron) under Ar atmosphere. With this method, the graphene surfaces consist of a mixture of monolayer (ML) and bilayer (BL) graphene, with ML coverage of at least 70%. The quality, composition, homogeneity, and precise thickness of the graphene was first evaluated by atomic force microscopy (AFM) and Raman spectroscopy. Typical morphology of the graphene samples used for this study are reported in Fig. S1 in the Supplementary Information.

All experiments reported here were conducted under ultra-high vacuum (UHV) conditions, maintaining a base pressure better than $5 \times 10^{-11}$ mbar. Prior to Pt deposition, the graphene samples were annealed, via direct current heating, at 600°C for several hours, followed by 10 minutes at 800°C to eliminate adsorbents and achieve clean surfaces. All temperatures were measured using a thermocouple mounted on the sample holder, directly in contact with the sample, and cross-calibrated with a pyrometer. The high quality of the pristine graphene films was confirmed by atomically resolved scanning tunneling microscopy (STM) images. The scanning tunneling microscope employed for these experiments is a VT-RHK-STM, working in constant current mode. Gwyddion software package was used to analyze the STM images [33].

Platinum deposition on graphene was carried out at room temperature using a commercial electron-beam evaporator, with the Pt coverage calibrated via STM imaging.

Hydrogenation of Pt clusters on graphene for thermal desorption spectroscopy (TDS) measurements was accomplished by exposing them to molecular deuterium for 5 min at a pressure of $1 \times 10^{-7}$ mbar at 25°C. Deuterium ($D_2$, mass 4) is chemically identical to hydrogen ($H_2$, mass 2). Still, it is less abundant in the residual atmosphere of the vacuum chamber, thus leading to a better signal-to-noise ratio in TDS. For the TDS measurements, the samples were positioned in front of a mass spectrometer (SRS-RGA) and heated at a constant rate of ~10°C/s to the target temperature (either 400°C or 750°C), while recording the mass 4-channel of the mass spectrometer.

## 3. Results and Discussion

First, we analyzed the Pt growth on graphene. In Fig. 1(a) we show the Pt area coverage (i.e., the amount of graphene surface area covered by Pt clusters with respect to the total area scanned, namely (100 nm × 100 nm)) versus the total volume of the Pt clusters, upon room temperature Pt deposition at constant flux. Herein, 1 ML is defined as the Pt atom density in Pt(111). A lattice constant of 0.392 nm yields 1 ML = $1.52 \times 10^{15}$ atoms/cm2 [34]. Whereas the interlayer distance is about 0.23 nm. Figure 1(a)-(c) shows three STM images of the epitaxial graphene surface with 0.53 ML, 1.39 ML, and 5.4 ML of Pt. These scans are representative of the surface morphology in the Pt growth regimes that we identified and which will be discussed in the following.

Up to a Pt coverage of about 1 ML, Pt particles tend to randomly disperse on the graphene surface. The cluster density increases from just a few clusters to about 185 clusters per 100×100 nm2, while their average height remains essentially unchanged. A typical surface morphology in this regime is shown in Fig. 1(b). This scan corresponds to a surface with 0.53 ML of Pt covering 26% of the graphene area. The Pt clusters have diameters between 1.3 nm and 10 nm, and the resulting average diameter is ~4 nm, while the cluster height spreads from 0.4 nm to 2 nm, resulting in an average height of ~0.6 nm. In this growth regime, we observe a linear relation between Pt volume and area coverage with intercept zero.

Afterwards, for Pt coverages > 3 ML the clusters merge, and due to coalescence their density decreases from ~35 clusters up to a single cluster per 100×100 nm2, at full area coverage, while their average height increases remarkably. With a Pt content of 5.4 ML, thus 93% area coverage (as shown in Fig. 1 (d)), the average height increases to 1.5 nm, while the maximum height reaches 4.5 nm. Notably, also in this regime we observe a linear relation between Pt volume and area coverage but with a reduced slope and intercept at about 50% area coverage. For large volume, the curve asymptotically goes to 100% area coverage.

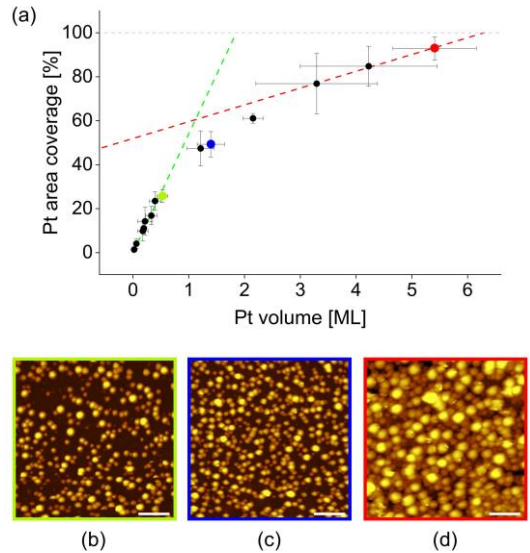

*Figure 1: (a) Area coverage [%] vs. Total cluster volume [ML]. All data points are the result of a statistical analysis of several STM images of size (100 nm × 100 nm). The error is calculated as the standard deviation for each dataset. Bottom: STM scans of (100 nm × 100 nm) areas representative of the surface with a Pt content (b) 0.53 ML - 26% (green), (c) 1.39 ML - 49% (blue), and (d) 5.4 ML - 93% (red). Colors for the frames of STM images (b)-(d) have been used accordingly to the colors of the data points highlighted in (a). Scale bar: 20 nm.*

These two regimes of growth kinetics can be better understood if we take into account the strength of interaction of Pt with the substrate (i.e., graphene) as opposed to Pt-Pt (the cohesive energy). It is important to emphasize that no Pt intercalation was observed during room temperature deposition. At low coverage, the growth of the clusters primarily involves competition between adsorption and desorption on the graphene surface. Additionally, if all evaporated Pt atoms adhered to the graphene surface and the film would grow with a layer-by-layer mechanism, we would have reached area coverage saturation at 1 ML. However, the linear fit for the first growth regime yields full coverage at ~1.85 ML, indicating that during the initial stages of growth the average cluster height is almost 2 layers, "approximately consistent" with the measured value of 0.6 nm. At higher coverage, the uncovered graphene surface area diminishes, and the dominant mechanism of growth shifts to nucleation of new layers on pre-existing clusters. Furthermore, since the flux of the evaporator is maintained constant, the non-zero intercept obtained from a linear fit of the second growth regime supports and implies that the Pt-graphene interaction strength is lower than that of Pt-Pt. This is indeed consistent with the reported values of Pt-Pt cohesive energy of 5.84 eV [35] which is more than two times larger than 2.16 eV, the computed binding energy of Pt on graphene [31].

We then analyzed the variation of hydrogen adsorption on Pt-covered graphene with increasing Pt content. For this measurement, we consistently evaporated increasing Pt amounts onto the sample. After each Pt deposition, we exposed the sample to molecular hydrogen (D2) and performed TDS. The sample was heated to 400°C, a temperature at which the shape and distribution of the Pt islands remained largely unchanged, as confirmed by STM analysis. Higher temperatures lead to significant and non-reversible morphological changes (refer to Fig. S2 in the Supplementary Information), thus the choice to stop TDS at ~400°C. Figure 2 reports the resulting temperature-dependent desorption curves of D2 as a function of Pt coverage.

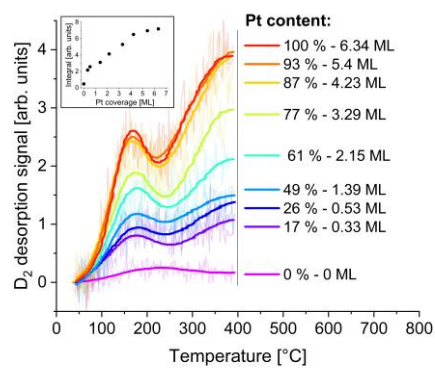

*Figure 2: TDS spectra of hydrogen desorption from Pt clusters supported on graphene. Inset: Integral intensity under the TDS spectra (from 25°C to 400°C) as a function of Pt coverage. Pt coverage ranging from 0 to 6.34 ML.*

Pure graphene exposed to molecular hydrogen showed little (see 0% in Fig. 2) to no D2 adsorption (refer to Fig. S1(d) in the Supplementary Information), consistent with the fact that molecular hydrogen does not stick to defect-free graphene at room temperature [6].

The D2 desorption spectra of Pt-clusters supported on graphene shown in Fig. 2, exhibit a peak at ~150°C followed by a broad shoulder of increasing signal. As the Pt coverage increases, the peak at 150°C in the desorption spectra becomes more pronounced, and the full integrated desorption signal continues to rise steadily, as demonstrated in the inset to Fig. 2. However, it eventually levels off around 87% of Pt area coverage.

For a more quantitative and detailed analysis of the desorption spectra of Pt-covered graphene, we performed a TDS experiment up to 750°C. For this purpose, we prepared a sample with a Pt coverage of 1.4 ML, corresponding to a Pt area coverage of 49% (see Fig. 1(c)). The TDS data is shown in Fig. 3.

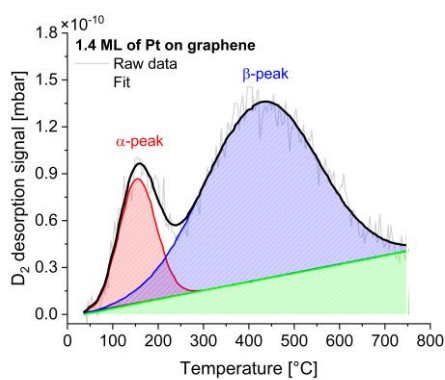

*Figure 3: TDS spectrum of hydrogen desorption from Pt clusters supported on graphene. Two Gaussian functions (red and blue curves) and a linear background (green curve) have been used for the fitting procedure.*

Comparing the first part of the TDS spectrum up to 400°C in Fig. 3 with the TDS spectra discussed above and presented in Fig. 2, we find consistent results between the two samples and can, again, identify a maximum in the desorption spectrum at ~150°C followed by a broad shoulder, which can now be resolved as an additional desorption peak.

The line profile analysis of this TDS spectrum actually reveals the presence of two Gaussian hydrogen desorption peaks labeled as α and β, superimposed on linearly increasing background. The α-peak is centred at $T_\alpha = (155\pm7)$°C, while the β-peak is centered at $T_\beta = (432\pm13)$°C. Assuming first-order desorption, we can use the following equation $E_d = k_B T_d = A\, \tau_d\, e^{-E_d} = k_B T_d$ [36] to evaluate an approximate desorption energy barrier ($E_d$ with d = α, β). Here, $\tau_d$ is the time from the start of the desorption ramp to the moment at which the desorption temperature $T_d$ is reached, and $k_B$ is the Boltzmann constant. We use for the Arrhenius constant (A) a typical value of $10^{13}$ s$^{-1}$. Then, we obtain $E_\alpha$ = 1.14 eV/molecule (0.57 eV/atom) and $E_\beta$ = 1.96 eV/molecule (0.98 eV/atom). Using the Readhead equation [37] we obtain similar results for the desorption barriers, $E_\alpha$ = 0.71 eV/atom and $E_\beta$ = 1.14 eV/atom.

The origin of these multiple hydrogen desorption peaks can be better understood comparing these data with literature. Molecular hydrogen can spontaneously dissociate to individual H atoms on the platinum surface [31, 32]. These H atoms can then chemisorb on the surface of the Pt clusters. Depending on the number of Pt atoms forming the active cluster and the Pt-support interaction, the strength of the bond between Pt and H atoms varies. The upper limit is set by an isolated Pt atom which can chemisorb up to 8 H atoms with adsorption energies in the range from -1.47 eV to -0.62 eV per H atom [31]. As these adsorption energies are reduced by the size of the cluster and the interaction of the cluster with the substrate [38], here epitaxial graphene grown on 6H-SiC(0001), they are in good agreement with the desorption energy we obtained for the β-peak. Thus, we can assign the β-peak to H atoms chemisorbed on the surface of the Pt clusters. The relatively large width of the peak is consistent with this interpretation.

Hydrogen can alternatively physisorb on the Pt cluster or chemisorb on the graphene surface by the spillover effect. TDS measurement on the $H_2$ adsorption on Pt(111) reported the occurrence of physisorption as single peak located at around 150-200°C [39, 40]. Chemisorption of atomic hydrogen on graphene has been reported to fall in the same energy range [41, 42]. For the latter to happen, the $H_2$ molecule dissociated by the Pt cluster has to leave the cluster and diffuse as an atom on the graphene substrate to form C-H bonds. According to theory, H can migrate from the metal cluster to the support only if this is thermodynamically more favorable. Theoretical works have reported activation energies for this process above 2 eV, thus concluding that this effect is unlikely to happen [31, 43, 44]. At the same time, there are other theoretical proposals and experimental reports that reveal the occurrence of spillover in TM-doped graphene [45-49]. In any case, since the spillover effect requires both an active catalyst and a support, the $H_2$ desorption peak associated to it must decrease to zero as the Pt reaches full coverage of the graphene surface. Since the α-peak does not disappear at full area coverage, therefore spillover is not likely to be the dominant mechanism responsible for this peak. Thus, we assign it to physisorbed hydrogen on the surface of Pt clusters.

Finally, for what concerns the linearly increasing background, we suggest that it is due to a "delayed effect" of desorption of hydrogen absorbed in the bulk of the Pt clusters [50]. This contribution should scale with the volume of the clusters, as the distance that the hydrogen has to travel before reaching the surface increases accordingly. As such, we should observe a significant increase of this contribution as the volume of the clusters becomes more important. From the TDS spectra presented in Fig. 2 we have an indication of an increasing shoulder which becomes more important at coverages higher than 75%. This is the same coverage at which we observed by STM an increased volumetric growth of the clusters and the onset of the second growth regime. However, since in the TDS spectra in Fig. 2 we observe a superposition of two Gaussian peaks with the background, the individual contributions are difficult to separate and thus our data cannot provide a conclusive demonstration of the origin of the background. Nevertheless, the interpretation fits well with what has been previously reported in literature [50, 51].

## 4. Conclusions

We presented an in-situ experiment on the growth and the hydrogen adsorption properties of Pt clusters supported on epitaxial graphene grown on 6H-SiC(0001).

As we increased the amount of deposited Pt, we monitored the cluster size and height distribution via scanning tunneling microscopy imaging. We identified two main growth modes which become competing at intermediate coverages (~2 ML). At low coverages (< 1 ML), Pt tends to adsorb and randomly disperse on the graphene surface and forms small clusters with an average height of ~0.6 nm. At higher coverages (> 3 ML), nucleation on pre-existing clusters becomes the dominant growth mechanism, and the cluster height tends to increase, reaching a maximum height of ~4.5 nm at 6.3 ML when full area coverage is reached.

With a deeper knowledge of the Pt dispersion on epitaxial graphene, we then studied via thermal desorption spectroscopy the hydrogen adsorption properties of the system. We show that the Pt-functionalized graphene system can adsorb hydrogen at temperatures that are high enough for stable binding at room temperature. The amount of stored hydrogen increases with the Pt content of the sample until it levels off at the onset of the second Pt growth regime. We discovered the presence of multiple hydrogen desorption processes, revealed as two Gaussian peaks (at about 155°C and 432°C) superimposed onto a linearly increasing background. From a comparison of the obtained TDS spectra with the data available in literature, we attribute the first Gaussian peak to hydrogen physisorbed on the surface of the Pt clusters. We attribute the second peak, instead, to dissociative chemisorption of the hydrogen molecules on the surface of the Pt clusters. On the other hand, the linear background is due to absorption of hydrogen in the bulk of the Pt clusters. Further studies are needed to better separate these contributions and to solve the controversial question whether hydrogen spillover can or cannot occur in this system.

## Acknowledgements

The research leading to these results has received founding from the European Union's Horizon 2020 research and innovation program under grant agreement no. 881603-GrapheneCore3.

# Supplementary Information

# Platinum-Decorated Graphene: Experimental Insight into Growth Mechanisms and Hydrogen Adsorption Properties

Letizia Ferbel[a,+], Stefano Veronesi[a], Ylea Vlamidis[a,b], Antonio Rossi[c,d], Leonardo Sabattini[c,e], Camilla Coletti[c,d], Stefan Heun[a]

[a]NEST, Istituto Nanoscienze-CNR and Scuola Normale Superiore, Piazza San Silvestro 12, Pisa, 56127, Italy
[b]Department of Physical Science, Earth and Environment, University of Siena, Via Roma 56, Siena, 53100, Italy
[c]Center for Nanotechnology Innovation@NEST, Istituto Italiano di Tecnologia, Piazza San Silvestro 12, Pisa, 56127, Italy
[d]Graphene Laboratories, Istituto Italiano di Tecnologia, Genova, 16163, Italy
[e]NEST, Scuola Normale Superiore, Piazza San Silvestro 12, Pisa, 56126, Italy
[+] letizia.ferbel@sns.it (Corresponding author)

## 5. Characterization of Pristine Epitaxial Graphene on 6H-SiC(0001)

The AFM used for the study is a Bruker Dimension icon operating in Peak Force QNM mode for nanomechanical mapping. AFM shows that the surface is characterized by terraces of width hundreds of nm to micrometer. Statistical analysis on the samples yields a coverage of monolayer graphene of at least 70%, while bilayer inclusion never exceeded 30%. No buffer layer areas were observed. Typical AFM scans are shown in Fig. S1(a)-(b).

Raman spectroscopy and mapping (20 × 20 nm2, with 400 pixels) was performed using a Renishaw inVia system with a 100× objective (NA 0.85) equipped with a 532 nm laser as excitation source. A typical averaged Raman spectrum is reported in Fig. S1(c).

TDS spectra of pristine epitaxial graphene obtained after sample exposure at $1\times10^{-7}$ mbar of $D_2$ for 5 min, how little to no hydrogen adsorption, as expected (as shown in Fig. S1d).

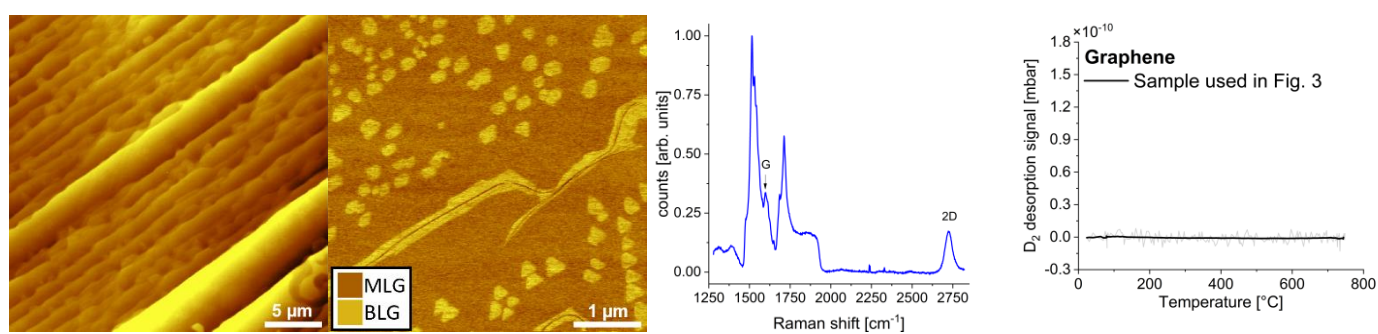

*Figure S3: Representative morphology and composition of the pristine epitaxial graphene samples used for this study. (a) AFM height image of a 30 μm × 30 μm area. (b) AFM adhesion image of a zoom-in 5 μm × 5 μm large. (c) Averaged Raman spectrum (20 μm × 20 μm, with 400 pixels). (d) TDS spectrum of hydrogen desorption from pristine epitaxial graphene.*

## 6. Effect of TDS on the sample morphology

TDS was performed with a ramp rate of ≃ 10°C/s. When the final temperature was reached the annealing was immediately stopped and the sample was left to cool down to RT.

The morphology of the sample does not significantly change after a TDS up to 400°C. The same cannot be said after a TDS up to 750°C. In this case, the area of graphene covered by the clusters reduces to about 40% while their average height more than doubles. The volume of the clusters results almost conserved.

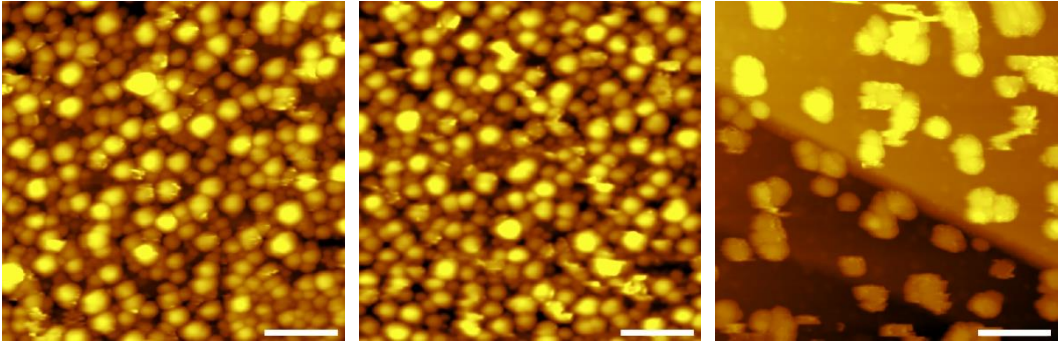

*Figure S2: STM scans of graphene covered by 5.4 ML of Pt. Scans taken (a) right after Pt deposition (Pt area coverage 93%), (b) after TDS up to 400°C (Pt area coverage 93%), and (c) after TDS up to 750°C (Pt area coverage ~40%). Scan area: 100 nm×100 nm. Scale bar: 20 nm.*